\documentclass[letterpaper,twocolumn,aps,prb,showpacs,floatfix]{revtex4-1}

\usepackage{graphicx}
\usepackage{subfigure}
\usepackage[latin1]{inputenc}
\usepackage{amsmath,amssymb,amsfonts}
\usepackage{amstext, mathrsfs, textcomp}
\usepackage{multirow}

\usepackage{hyperref}

\usepackage{xcolor}
\graphicspath{{FIGURES/}}

\begin{document}

\title{Enhanced nonlinear optical response from individual silicon nanowires}

\author{\firstname{Peter R.} \surname{Wiecha}$^{1,2}$}
\author{\firstname{Arnaud} \surname{Arbouet}$^{1}$}
\author{\firstname{Houssem} \surname{Kallel}$^{1,2}$}
\author{\firstname{Priyanka} \surname{Periwal}$^{3}$}
\author{\firstname{Thierry} \surname{Baron}$^{3}$}
\author{\firstname{Vincent} \surname{Paillard}$^{1,2}$}
\email[corresponding author~: ]{vincent.paillard@cemes.fr}

\affiliation{$^{1}$ CEMES-CNRS, 29 rue Jeanne Marvig, 31055 Toulouse cedex 4, France}
\affiliation{$^{2}$ $Universit\acute{e}$ de Toulouse, $Universit\acute{e}$ Paul Sabatier, 118 route de Narbonne, 31062 Toulouse cedex 9, France}
\affiliation{$^{3}$ LTM/CNRS-CEA-Universit$\acute{e}$ de Grenoble Alpes, 17 rue des martyrs, 38054 Grenoble, France}

\begin{abstract}
We report about the experimental observation and characterization of nonlinear optical properties of individual silicon nanowires of different dimensions. Our results show that the nonlinear light has different components, one of them corresponding to the second harmonic generation (SHG). The SHG strongly depends on the polarization of the optical excitation and nanowire diameter, and gives access to the  local electromagnetic field intensity distribution. Furthermore, we show that the second harmonic, when observed, is enhanced compared to bulk silicon and is sensitive to optical resonances supported by the nanowires. This offers different perspectives on the definition of silicon-based nonlinear photonic devices.
\end{abstract}


\maketitle

High-refractive index dielectric nanostructures provide original optical properties, exhibiting intense resonances depending on their size, shape and the constituent materials. In particular, semiconductor nanowires can sustain so-called leaky mode resonances (LMRs), which depend on the material and the perimeter.\cite{Cao10,Bro14,Kal12} These morphological resonances occur when the perimeter is commensurate with the wavelength inside the nanowire. 
Similarly, guided modes can be excited and different Fabry-Perot resonances appear as function of the nanowire length.\cite{Wel12}

Though any semiconductor or insulating material with high refractive index can be used to engineer light-matter interaction and to design efficient optical resonators for light trapping, silicon nanostructures such as nanowires (Si-NWs) are particularly promising for low cost photonic devices compatible with CMOS technology.\cite{Bro14,Pri14}
It has been shown that Si nanowires could be used as antennas on top of a thin planar solar cell.\cite{Bro14,Fer11} The key parameter is the enhanced scattering efficiency of the nanostructure and the optical interaction with the thin film where the electron-hole pair generation occurs. The amplified electromagnetic field should thus enable an increased photon absorption in the solar cell active layer.

Si-NWs can be used also for field-enhanced spectroscopy as they present interesting near-field enhancements but, more importantly, much lower losses compared to plasmonic nanostructures.\cite{Mai14} Guided modes\cite{Wel12} or LMRs\cite{Pan13} have been used to enhance Raman scattering\cite{Wel12} or photoluminescence from molecules or nanocrystals,\cite{Wel12,Kal13} 

For these applications, it is important to have access to the optical near-field associated with the leaky- or guided modes.\cite{Pan13}  Aside from optical near-field  experiments, complex in both their use and interpretation, several alternative far-field techniques such as nonlinear optical microscopies have been developed to obtain direct or indirect information on the near-field distribution, for instance in the case of III-V semiconductor nanowires \cite{Gra12,San14} or noble metal nanostructures.\cite{Ghe08,Ber12}

In the following, we investigate  individual Si-NWs of different diameters using nonlinear optical microscopy. We show that an intense SHG is emitted, which is dependent on the incident light polarization. We exploit the SHG, which is highly sensitive to small variations of the local electric field and boosts the resolution of the SHG mappings of a NW, to access to the local electromagnetic (EM) field distribution of isolated Si-NWs.
We demonstrate that the SHG is influenced by the dielectric contrast between the NW and its environment, and the existence of resonant optical modes supported by the NW. We prove that such modes lead to an enhanced SHG yield of Si-NWs, especially compared to bulk silicon.

The growth of Si-NWs of different diameters and the production of isolated Si-NWs on glass substrates is described elsewhere. \cite{Kal12,Kal13} Among all investigated samples, we selected three nanowires of 47, 80 and 295 nm diameters and at least 2 \textmu m in length, referred to as NW50, NW80 and NW300, respectively. The smallest diameters were chosen because they support a limited number of LMRs (0 to 2 depending on the polarization). Larger nanowires have a roughly similar behavior.

Non-linear optical microscopy was performed using a homemade setup described in Fig.~\ref{fig1}a. Details are given in Supplemental Information.
{\color{red}(Section I: Nonlinear optical microscopy experiments)}
Power dependence measurements show that the detected light is a consequence of nonlinear processes, but its behavior completely changes as function of the power density, polarization and position on the nanowire (tip, center) of the incident laser beam.

\begin{figure}[tb]
\centering
\includegraphics*{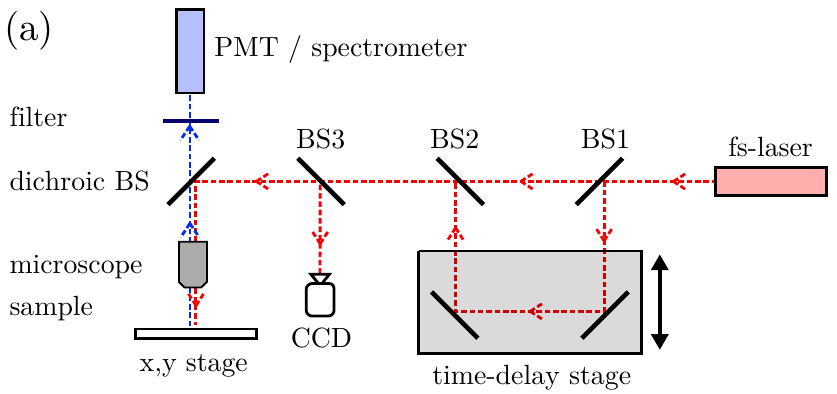} 
\includegraphics*{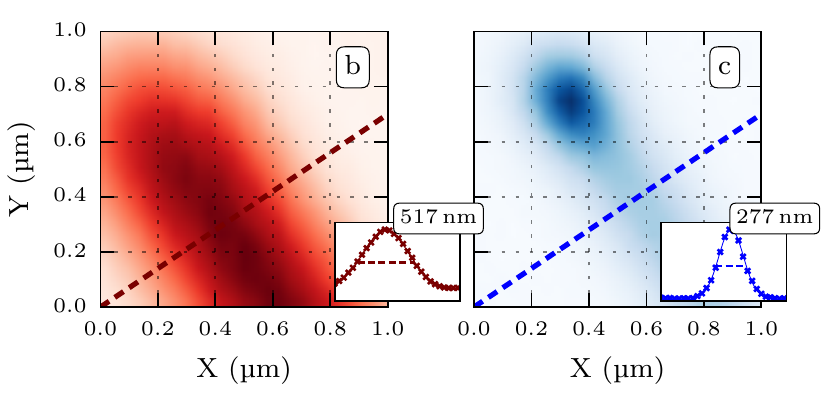} 
\caption{a) Schematic view of the set-up used for SHG measurements. The second beam passing through the time-delay line is used for interferometric autocorrelation experiments only. (BS: beamsplitter), b) Elastic light scattering mapping of an isolated Si-NW, c) Nonlinear emission mapping. In insets are shown the FWHM taken long the dashed line}
\label{fig1}
\end{figure}

\begin{figure}[tb]
\centering
\includegraphics*{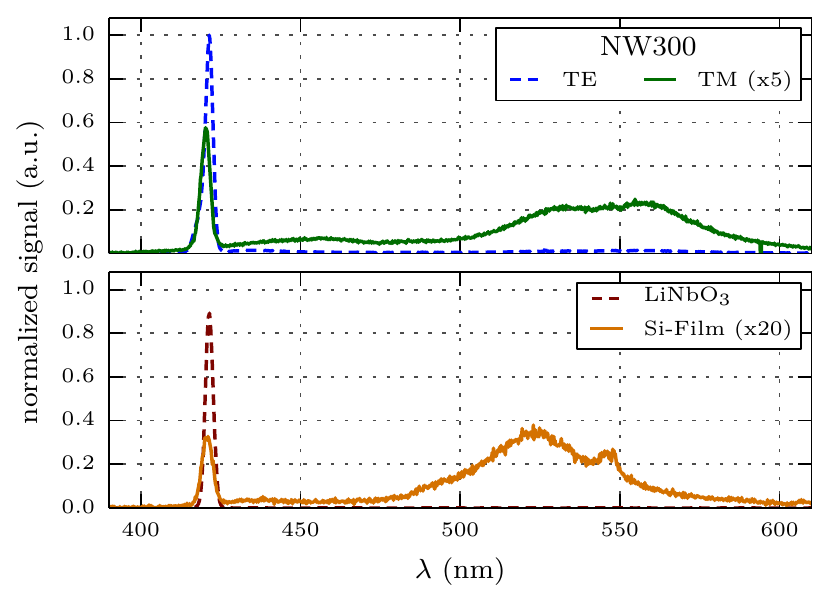} 
\caption{Spectra of SHG and nonlinear photoluminescence of NW300 in TE  and TM configurations with excitation at $\lambda=840\,$nm on the center of the NW. The laser polarization was fixed while the samples were rotated by 90\,$^{\circ}$. The SHG signals of LiNbO$_3$ , and a 200\,nm thick Si layer of a Silicon on Quartz are shown for comparison. The average power density was 3.84\,mW/\textmu m$^2$ for Si materials and 0.92\,mW/\textmu m$^2$ for LiNbO$_3$}
\label{fig2}
\end{figure}

Fig.~\ref{fig2} shows the spectrum of the nonlinear emission recorded at the center of a large diameter Si-NW such as NW300.  It is dependent on both the polarization and power density of the incident laser beam. The incident light is polarized parallel to the NW axis in TM configuration, and perpendicular to the NW axis in TE configuration.
At low power densities, the spectra show mainly a sharp peak that should be attributed to SHG, the peak wavelength being found at half the wavelength of the laser. A broad PL band ranging from the SHG wavelength to about 600\,nm (cut-off due to the dichroic mirror) is also present in TM configuration. With increasing power densities, this PL band can be detected for both TE and TM polarizations. 
These broad bands also exhibit a non-linear power dependence with a different exponent than the SH peak. Finally, the emission spectrum is dependent on the position of the laser spot on the NW (tip or center), in agreement with the assumption of local EM field sensitivity.

To unambiguously ascribe the sharp peaks in Fig.~\ref{fig2} to SHG and to investigate the origin of the other spectral features, power dependence and two pump interferometric autocorrelation measurements were performed. From the results and figures given in Supplemental Information, 
{\color{red}(Section I: Nonlinear optical microscopy experiments)}
we can conclude that the sharp peak corresponds to SHG, while a nonlinear order of 3 is found for the PL bands. 
We suggest that the third-order nonlinear PL bands may arise from the native oxide surrounding the NW. Such signal is neither detected from the glass substrate outside the NW, nor under continuous wave excitation at any position on or outside the NW. 
We believe that this luminescence could be generated by excitation of colored centers in silica such as self-trapped excitons, oxygen vacancies and H-related defects\cite{Rou11} by either Third Harmonic Generation in the Si-NW core, or by three photon absorption in the SiO$_{2}$ sheath. Similar spectral signatures have also been detected from a Silicon on Quartz (SOQ) sample formed by a 200 nm thick single crystal Si layer transferred on a fused silica substrate.

\begin{figure}[tb]
\centering
\includegraphics*{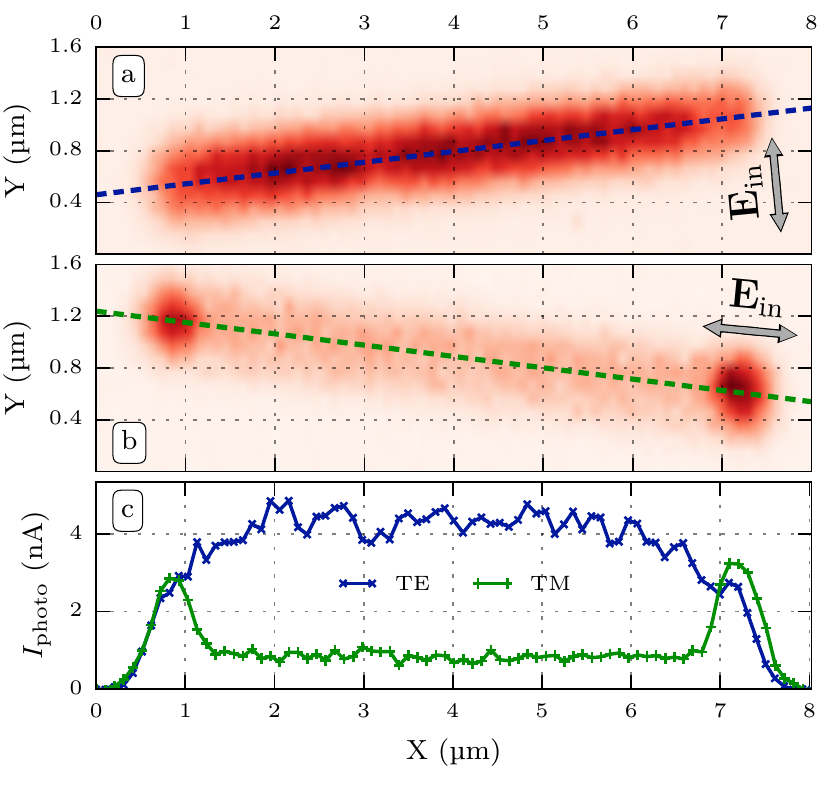} 
\caption{SHG maps of NW300 for (a) TE polarization, (b) TM polarization and (c) intensity profiles along the NW axis.}
\label{fig3}
\end{figure}

\begin{figure}[tb]
\centering
\includegraphics*{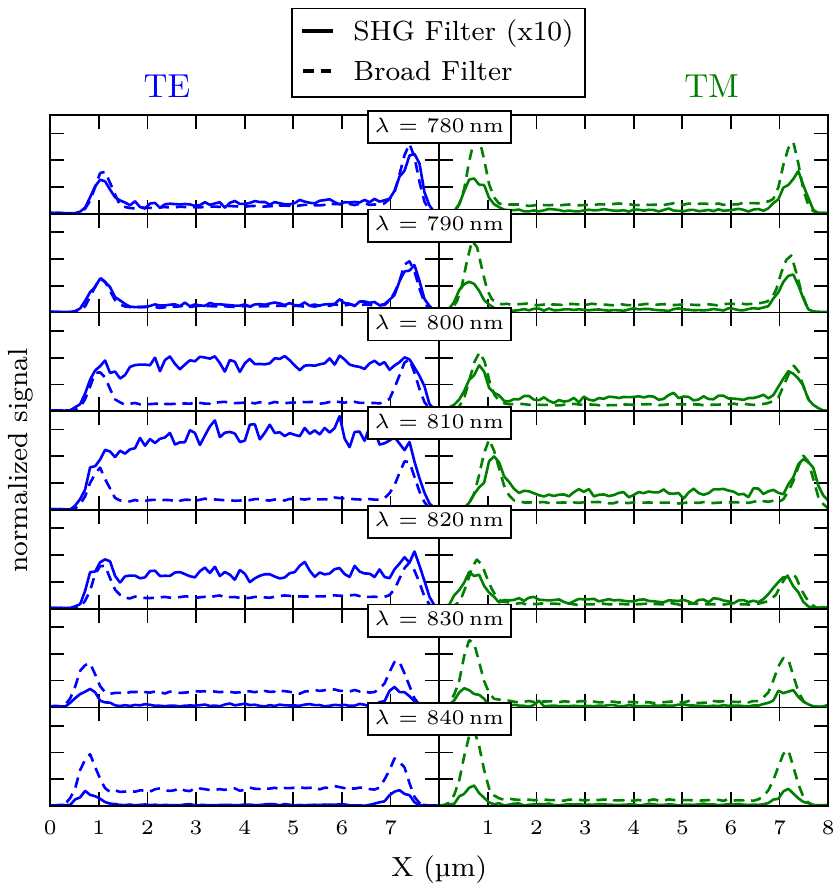}
\caption{Intensity profiles along the NW300 axis as function of the excitation wavelength for two different detection windows (SHG filter (solid line) and PL filter (dashed line)), and for both polarization configurations (TE blue, TM green).}
\label{fig4}
\end{figure}

Typical SHG maps recorded for both TM and TE configurations on large nanowires (diameter over 100\,nm) are shown in Fig.~\ref{fig3} and Fig.\ref{fig4}. The main difference is that SHG is homogeneous along the NW in the TE configuration, while it is strongly enhanced at both ends in the TM case. This behavior suggests that one of the contributions of SHG in Si-NWs is a surface-induced process like in bulk silicon, where the lattice symmetry is broken by the presence of surfaces or interfaces.\cite{Tom83,Kau05} Another cause of symmetry breaking for SHG in Si is the introduction of strain,\cite{Caz12} which is not the case of our NWs.\cite{Raman}

The strong anisotropy between TE and TM maps of SHG can be explained by the continuity relations for the EM field at the interface between two media presenting a large dielectric contrast.\cite{Bar08} The relations between external and internal fields show that the external field is constant when it is parallel to the NW surface and is increased when it is normal to the NW-air interface.\cite{Cis11}
In the TE configuration case, the incident field is always perpendicular to the NW axis. Therefore, the external near-field is homogeneously enhanced all along the NW, in agreement with the SHG mapping of Fig.~\ref{fig3}a. 
In the TM case, the incident field has only a tangential component with respect to the surface along the NW length, except at both NW edges where it is normal to the termination surface. The field enhancement is thus localized at the NW tips, as the SHG enhancement as shown in Fig.~\ref{fig3}b.

%

Another evidence of the SHG sensitivity to the dielectric contrast is given in Fig. \ref{fig4}, where the intensity profiles are drawn along the NW300 axis for different excitation wavelengths at a fixed detection window (SHG filter). The difference between the TM and TE configurations is detected only when the excitation wavelength is equal to twice the detection wavelength. In the other cases, as well as for the profiles recorded using the longpass filter (corresponding to the broad PL band induced by third order processes), a totally different behavior is obtained. All profiles except the SHG ones look very similar, are independent of the incident light polarization, with "hot spots" located at both NW tips.

There are different sources of SHG in centrosymmetric materials. \cite{Cis11,Kau05,Bac10} These sources are divided between surface local contributions due to the rapid change of the electric field at the interface between the nano-object and its embedding medium and non-local bulk contributions which depend on the gradient of the incident electric field.
It is generally assumed that the normal non-linear surface polarization is the main contribution to SHG from gold nanostructures, this contribution being highly sensitive to the shape of the nano-object,\cite{But13a} or proximity to another nano-object.\cite{But13b}
In the case of silicon nanostructures, the relative weight of surface and bulk contributions is not well-known. Extensive numerical simulations are required to address the SHG mechanism from Si-NWs. 

The first conclusion of our work is that far-field SHG microscopy performed on silicon nanowires gives access to the local near-field intensity distribution at the fundamental frequency, in agreement with similar experiments on gold nanorods.\cite{Ber12} The comparison of the SHG maps of Fig.~\ref{fig3} and the results of electrodynamical simulations provided in supplementary information indeed shows that the general aspect, TE and TM polarization dependence and intensity ratio are qualitatively recovered in our experiments.
{\color{red}(See Supplemental Information, Section II: Near-field intensity distribution of a silicon nanowire)}
This conclusion is supported by previous study of the photoluminescence (PL) of silicon nanocrystals located in the optical near-field of Si-NWs, \cite{Kal13} which showed that the PL mapping was in good agreement with the computed near-field intensity distribution of the Si-NW.\cite{Kal13}

At last, it must be noticed that SH detected in our experiments can be strongly enhanced in Si-NWs compared to bulk silicon. The values of the SHG yield for different materials excited under the same conditions are summarized in Table~\ref{tab1}. The measured SHG intensities are higher for Si-NWs than for bulk Si and SOQ layer, even more when the SHG signal is renormalized by the area $K_{\text{spot}}$ of the laser spot that is actually illuminating the Si-NW. 
{\color{red}(See Supplemental Information, Section III: Correction factor for the SHG of silicon nanowires and bulk materials)}
This correction is necessary since the minimal laser spot radius ($\approx\! 620\,$nm for $\lambda_{\text{exc}}\!=\!810\,$nm) is much larger than the Si-NW diameters. The corrected SHG signal of Si-NWs may be two orders of magnitude larger than bulk Si and SOQ layer.

\begin{table*}[tb]
\caption{Comparison of SHG signals. $I_{\text{photo}}$/$K_{\text{spot}}$ is the normalized photocurrent (see text). Laser power and wavelength were fixed at 3.15\,mW/\textmu m$^2$ and 810\,nm, respectively. The scattering efficiency  $Q_{\text{scat.}}$ at 810\,nm is calculated using Mie theory.\cite{Kal12}}
\begin{ruledtabular}
    \begin{tabular}{l||c|c|c|c|c|c}
      Sample          & Polarization    & Position 
                     & $I_{\text{photo}}$ (nA) & $K_{\text{spot}}$ & $I_{\text{photo}}$/$K_{\text{spot}}$ (nA) & $Q_{\text{scat.}}$  \\
      \hline \hline

         Glass    & --     & --        & 0.015    & 1           & 0.015     & --      \\                                                              
         \hline                                                                         
      Si Bulk                   & --      & --           & 0.19          & 1      & 0.19     & --        \\         
        \hline                                           
     SOQ      & --          & --            & 0.22                & 1       & 0.22      & --        \\
        \hline
      \multirow{2}{*}{NW50}     & TE     & Tip/Center    & 0/0    & \multirow{2}{*}{0.073}    & 0/0    & 0.014 \\
                                     
                                                & TM   & Tip/Center    &  2.5/1.6    &        & 34.2/21.9      & 3.32        \\
                                    
      \hline                                                                                                                          
      \multirow{2}{*}{NW80}    & TE    &    Tip/Center       &   0.30/0.30   & \multirow{2}{*}{0.116}   & 2.6/2.6  & 0.061  \\
                                              & TM      & Tip/Center    &   4.8/4.4   &        & 41.4/37.9          & 6.34    \\
                                   
      \hline                                                                                                                          
      \multirow{2}{*}{NW300}  & TE    & Tip/Center    & 3.3/3.8    & \multirow{2}{*}{0.436}  & 7.6/8.7    & 0.60 \\
                             
                                                 & TM      & Tip/Center       & 5.6/1.6    &     & 12.8/3.7         & 1.61       \\

\end{tabular}
\end{ruledtabular}
\label{tab1}
\end{table*}

The SHG yield is very different depending on the TM or TE configuration, and on the NW diameter. Therefore, it is interesting to compare the $I_{\text{photo}}$/$K_{\text{spot}}$ results in view of the scattering efficiency at the excitation wavelength, thus addressing another contribution to the SHG yield than surface and dielectric contrast effects. Notice however that we did not take into account the optical absorption, since it is the same for all silicon samples (no resonance around 810\, nm for Si-NWs).

\begin{figure}[tb]
\centering
\includegraphics*{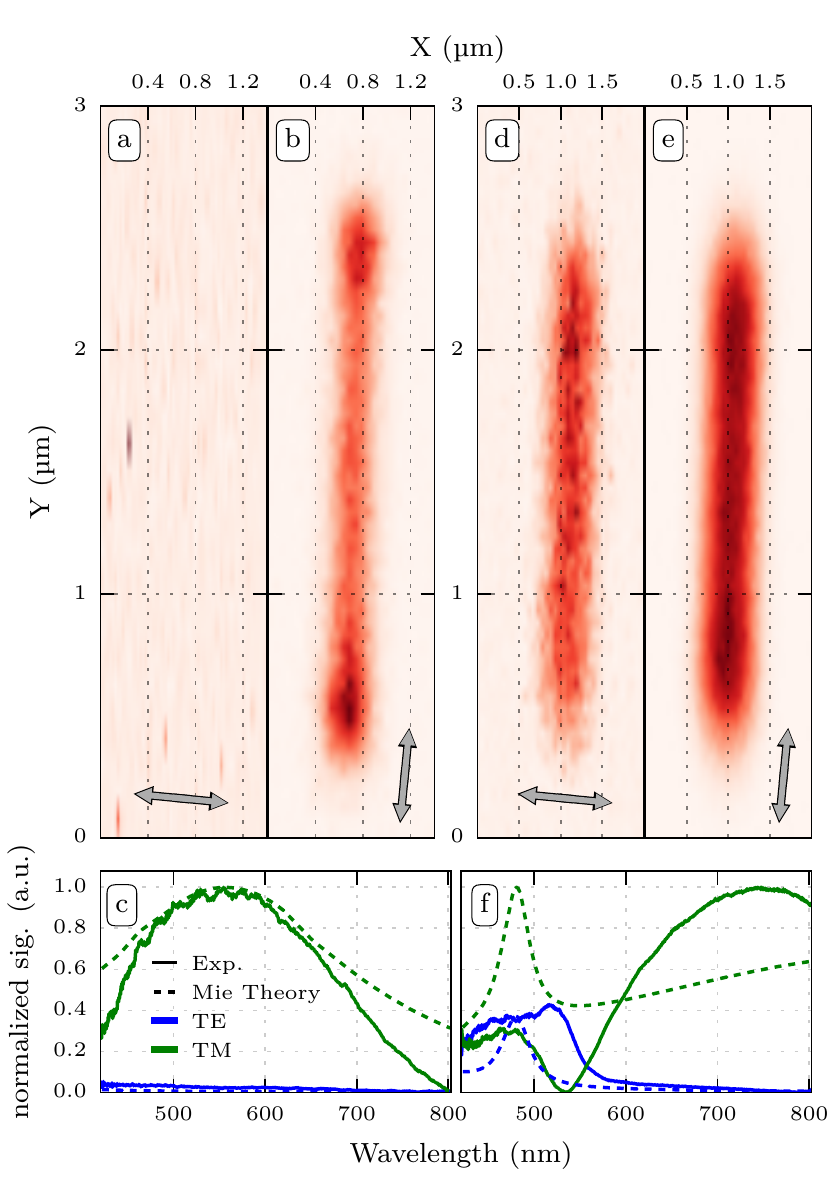}
\caption{SHG maps for (a) (resp. (d)) TE, and (b) (resp. (e)) TM polarizations in the case of NW50 (resp. NW80). Normalized elastic light scattering spectra compared to Mie theory for (e) NW50, and (f) NW80.}
\label{fig5}
\end{figure}

The impact of Mie resonances can be seen on the SHG mappings of small diameter NWs, such as NW50, supporting only one of these optical resonances (Fig.~\ref{fig5}a-c). In that case, the fundamental resonance branch is non-degenerate and only the TM$_{01}$ mode can be excited.\cite{Cao10,Kal12}  Thus, a high SHG yield is measured in the TM configuration, corresponding to a large scattering efficiency ($Q_{\text{sca}}\!=\!3.3$), while no SHG is detected for perpendicularly polarized incidence, due to the absence of any TE mode ($Q_{\text{sca}}\!=\!0$).
The behavior for NW80 (in Fig.~\ref{fig5}d-f) is quite similar to NW50, with a non-degenerate TM$_{01}$ mode matching the excitation wavelength (810\,nm). The higher SHG yield of NW80 in the TM configuration can be attributed to the large scattering efficiency at the excitation wavelength, and possibly at the detection wavelength (405\,nm). The SH intensity involving local field enhancement factors at both the fundamental and harmonic wavelengths has been reported in the case of gold nanorods.\cite{Hub07,Kau12}  However, a very weak SHG instead of no SHG for NW50 can be distinguished in the mapping  of the TE configuration (Fig.~\ref{fig5}d). This could be related to the value of $Q_{\text{scat}}$ at the excitation wavelength for NW80 which is slightly higher than $Q_{\text{scat}}$ for NW50 (see table~\ref{tab1}), and again to the presence of a resonance mode at the harmonic wavelength (degenerate TM$_{11}$-TE$_{01}$ mode around 450\,nm).
As mentioned previously, the influence of optical absorption can be neglected as the absorption efficiency is very weak for all NWs at the fundamental wavelength (even for large diameter Si-NWs above 200 nm where absorption resonances may appear).

In conclusion, we have studied the nonlinear optical response from individual Si-NWs of different diameters. We identified different nonlinear contributions, one of them being second harmonic generation. First, we exploited the SHG, which is highly sensitive to small variations of the local electric field and furthermore boosts the resolution of the SHG mappings of a nanowire, to access to the local electromagnetic field distribution of a Si-NW. Second, we showed that Si-NWs are good candidates for enhanced SHG, thanks to the presence of resonant optical modes. The optimization of the SHG yield could be achieved by adjusting the nanowire diameter (hence the resonances), for instance to match simultaneously the fundamental and harmonic frequencies. This may open a route towards new Si-based nonlinear optical devices. \\

The authors thank Dr C. Girard (CEMES) for fruitful discussions and Dr F. Fournel (CEA-LETI) for providing the SOQ substrate. The optical set-up was partly funded under "Campus Gaston Dupouy" grant. This work was supported by the computing facility center CALMIP under grant P12167.

\newpage
\widetext
\begin{center}
\textbf{\large Supplemental Materials: Enhanced nonlinear optical response from individual silicon nanowires}
\end{center}
\setcounter{equation}{0}
\setcounter{figure}{0}
\setcounter{table}{0}
\setcounter{page}{1}
\makeatletter
\renewcommand{\theequation}{S\arabic{equation}}
\renewcommand{\thefigure}{S\arabic{figure}}
\renewcommand{\bibnumfmt}[1]{[S#1]}
\renewcommand{\citenumfont}[1]{S#1}

\section{Nonlinear optical microscopy experiments}

The beam of a mode-locked Ti-sapphire femtosecond-laser covering the 700\,nm-1000\,nm wavelength range (80\,MHz repetition rate, 150\,fs pulse duration) is focused through a 50x, 0.8\,NA microscope objective. The emitted light is collected in the backscattering geometry using the same microscope. 

Spectra at different points of the samples are obtained using a Princeton Instruments IsoPlane 320 spectrometer coupled to a Pixies 400 CCD device. A nano-displacement stage is used to move the sample for 2D mapping (usually steps of 100\,nm). In that case, either longpass or bandpass filters and a photomultiplier tube (PMT) are used. The bandpass filter (FBH405, 405\,nm, 10\,nm FWHM) is used to select the second harmonic generated by $\lambda=810$\,nm laser light (the so-called SHG filter in the article). The longpass filter (GG435, starting at 435\,nm), or broad filter, is used to measure the nonlinear broad band emission generated by two or more absorbed photons. Finally, a BG39 bandpass filter (360-580\,nm) is always added to eliminate any unwanted fundamental light, leaking through the dichroic mirror.

The incident polarization is fixed. The TE and TM configurations (polarization perpendicular and parallel to the NW axis, respectively) are obtained by rotating the sample. The incident power density is thus never changed. Except for power dependence measurements, the average power density was fixed at 3.15\,mW/\textmu m$^2$. 

Interferometric autocorrelation experiments are performed using two coherent pumps.\cite{Mon10} The laser pulse is split in two beams of equal power, which are time-delayed before being focused on the sample in the same spot through the microscope objective. The average power density for each beam was fixed at 1.35\,mW/\textmu m$^2$.\\

\begin{figure}[h!]
\centering
\includegraphics*{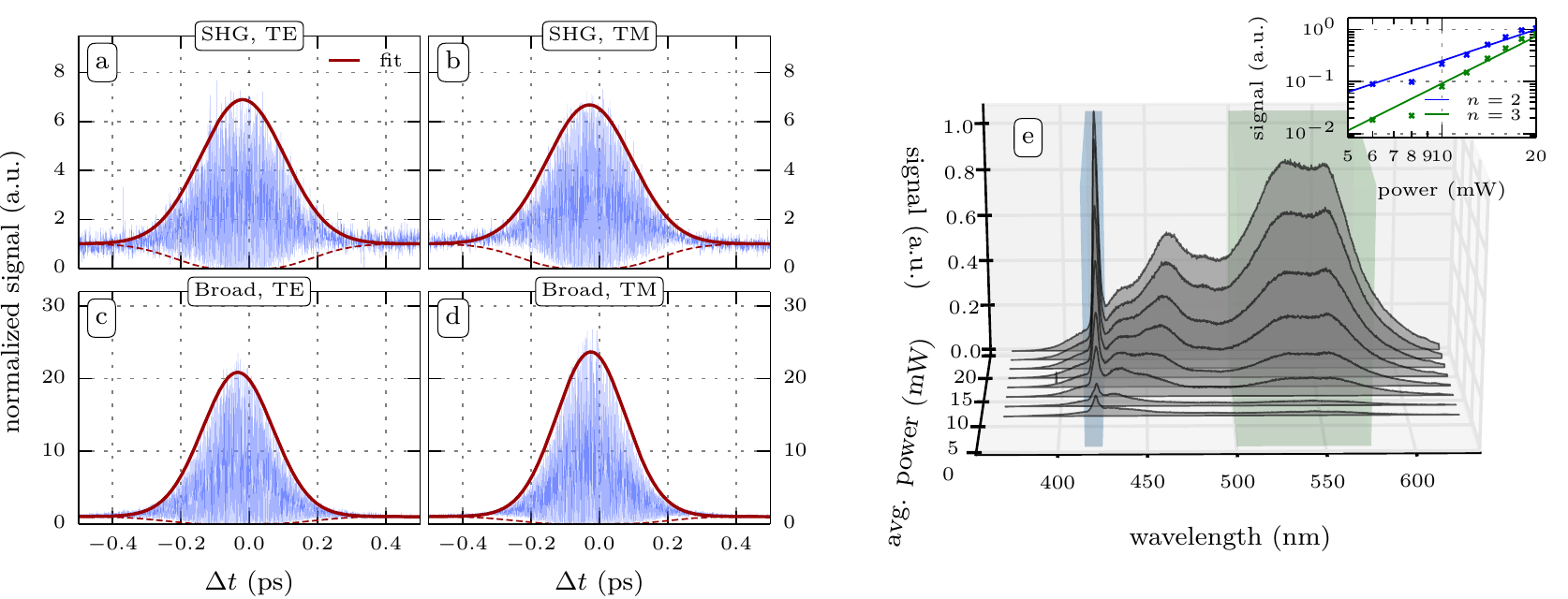}
\caption{Interferometric autocorrelation spectra for SHG ((a) and (b)) and broad nonlinear photoluminescence band ((c) and (d)). (e) Nonlinear spectra as function of laser power recorded on the NW tip in TM configuration. Log-log plots showing the power dependence are given in the inset ($n=2$ and $n=3$ slopes are drawn as a guide for the eye).}
\label{fig6}
\end{figure}

Power dependence measurements are shown in Fig.~\ref{fig6}e, where the spectra are given as function of increasing laser power. The log-log inset shows a slope of 2 for the sharp peak's maximum intensity (SHG), whereas the broad photoluminescence band rises with a slope of 3. For these experiments the fundamental light was 840\,nm (harmonic light at 420\,nm) to optimize the transmission of the microscope objective and the spectrometer's detection efficiency. Results were reproducible with increasing and then decreasing power density, as long as keeping the power density below a damage threshold.

To validate the use of filters for SHG mapping, we performed interferometric autocorrelation measurements on the SHG and the broad PL band.
The experiments were performed using a fundamental light of 810\,nm. Detection was limited to either second harmonic using a FBH405-bandpass filter or to the broad luminescence band using a GG435-longpass filter. Results are shown in Fig.~\ref{fig6}a-d.

The measured intensity of the interferometric autocorrelation signal at time delay $\delta$ writes:
%
\begin{equation}
I_{\text{NL}}(\delta) = \int\limits_{-\infty}^{\infty}  \left|\left[
E_{\text{pulse1}}(t) + E_{\text{pulse2}}(t-\delta) \right]^2\right|^n
\text{d}t
\end {equation}
with $E_{\text{pulse1,2}}(t)$ the electric field of pulse 1,2 at the sample at time $t$, and \textit{n}  the nonlinear exponent. Setting both pulses to equal amplitudes $A_{1}=A_{2}=A$, the ratio of maximum signal ($\delta=0$) to infinite delay signal is:
%
\begin{equation}
R =   \frac{I_{\text{NL}}(0) }{ I_{\text{NL}}(\infty)} = \frac{\left|[2
\cdot A]^2\right|^n}{2\cdot \left|A^2\right|^n} = \frac{4^n}{2}
\end {equation}
which gives $R=8$ for $n = 2$, $R=32$ for $n = 3$, ...

The fit to the envelope of the experimental spectra shows that the amplitude is nearly 8 for the spectrum recorded using the SHG filter (Fig.~\ref{fig6}(a) and (b)), thus corresponding to second harmonic generation. The amplitude for the large PL band is slightly inferior to 32, in agreement with a third order process (Fig.~\ref{fig6}(c) and (d)).

\section{Near-field intensity distribution of a silicon nanowire}
Fig.~\ref{fig8} shows the near-field intensity distribution at 10\,nm (one discretization step) below a Si-NW placed in vacuum calculated using the Green Dyadic technique.\cite{Gir05}  The latter relies on a volume discretization of the system and allows to describe nanostructures of arbitrary morphology. This technique rigorously takes into account the dielectric response of the substrate. Due to excessive computational requirements, it is impossible to perform a simulation of a 300\,nm diameter, 6\,\textmu m long Si nanowire. We limit the nanowire sizes to 2\,\textmu m in length and to 50 and 80\,nm in diameter, corresponding to the other sizes investigated experimentally. The near-field maps are in qualitative agreement with the SHG maps of Fig.~3 and Fig.~5 of the article. The electric field is strongly enhanced in TM configuration at the NW edges only. It is not or weakly enhanced but homogeneous all along the NW axis in the TE configuration. The absence or weak field enhancement in the TE case is due to the absence of any resonance for the small diameter Si-NWs at the excitation wavelength (810\,nm).

\begin{figure}[h!]
\centering
\includegraphics*{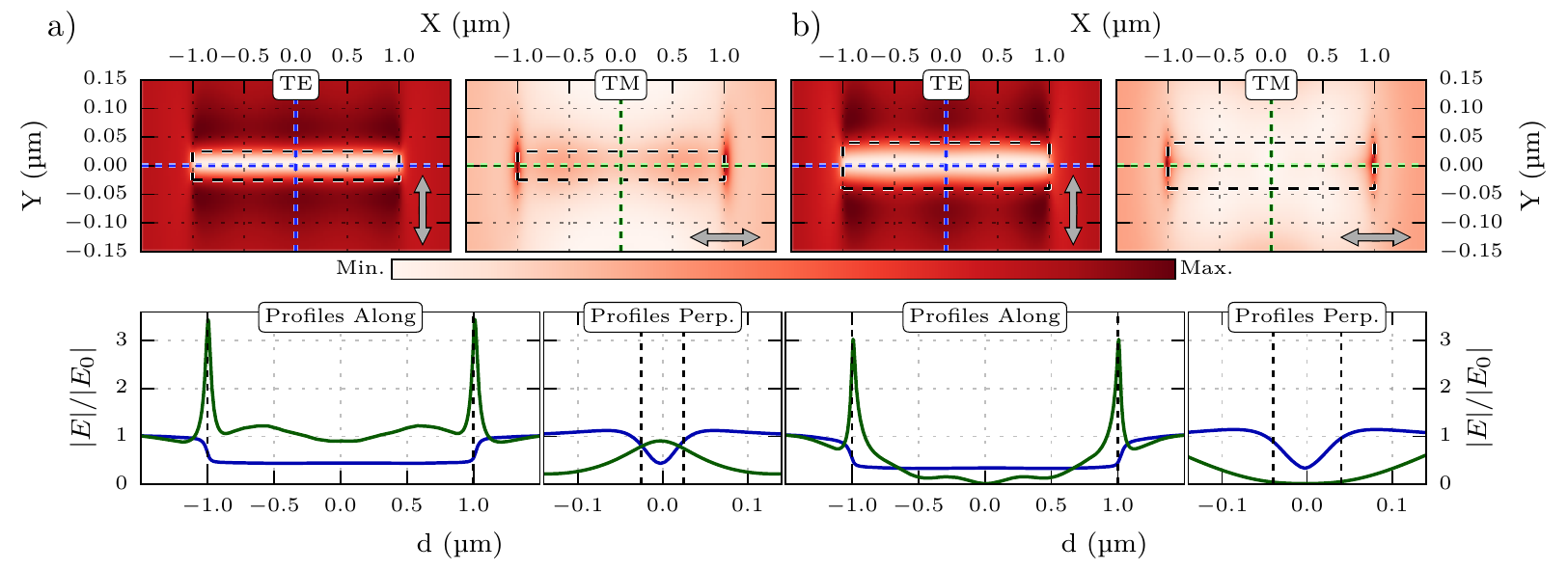}
\caption{Near-field intensity distribution of a 2\,\textmu m long Si-NW (diameter of 50\,nm (a) and 80\,nm (b)) for TE and TM configurations under plane wave excitation at 810\,nm. Lower figures : normalized intensity distribution drawn along the dotted blue and green lines in the upper images. In each case, profiles parallel and perpendicular to the NW axis are given.}
\label{fig7}
\end{figure}

\section{Correction factor for the SHG of silicon nanowires and bulk materials}

In our experimental conditions, the diffraction limited minimal spot radius $r_{\text{spot}}\!\approx\!0.61\lambda/\text{NA}\!\approx\! 620\,$nm (for $\lambda\!=\!810\,$nm) is always larger than the Si-NW diameter.
The illuminated fraction of the nanowire is estimated by a factor 
%
\begin{equation}
K_{\text{spot}} = \frac{S_{\text{onNW}}}{S_{\text{spot}}}
\end {equation}
with the intersection of spot and NW surface $S_{\text{onNW}}\!=\!d_{\text{NW}}\cdot 2r_{\text{spot}}$/$\sqrt2$, and \(S_{\text{spot}}\!=\!\pi (r_{\text{spot}}/\sqrt2)^2\). The spot area is taking into account the 2$^{\text{nd}}$ order of the nonlinear process (the factor 1/$\sqrt2$ is due to the squared Gaussian beam width).

The photocurrent is divided by $K_{\text{spot}}$ to obtain the normalized SHG yield, allowing the comparison of nanowires to bulk samples. The absorption efficiency at the excitation wavelength is the same for all silicon samples. Results are shown in Table II of the article.

\begin{figure}[h!]
\centering
\includegraphics*{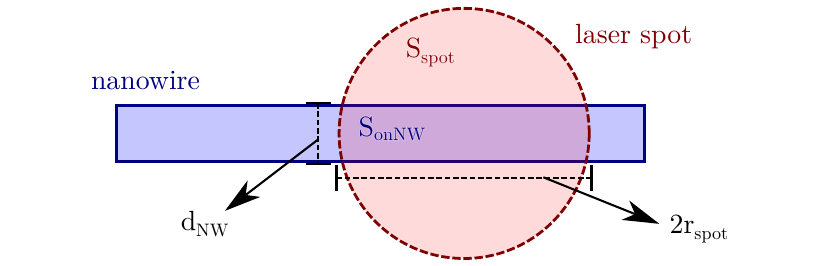}
\caption{Sketch showing the illuminated fraction of a nanowire by a circular laser spot.}
\label{fig8}
\end{figure}


\begin{thebibliography}{00}
\bibitem{Raman} Raman spectra of Si-NWs do not exhibit any significant strain-induced phonon shift.
\end{thebibliography}
\end{document}